\def\year{2022}\relax
\documentclass[letterpaper]{article} 
\usepackage{aaai22}  
\usepackage{times}  
\usepackage{helvet}  
\usepackage{courier}  
\usepackage[hyphens]{url}  
\usepackage{graphicx} 
\usepackage{natbib}  
\usepackage{caption} 
\DeclareCaptionStyle{ruled}{labelfont=normalfont,labelsep=colon,strut=off} 
\usepackage{algorithm}
\usepackage{algorithmic}
\usepackage{amsfonts}
\usepackage{xcolor}
\usepackage{newfloat}
\usepackage{listings}
\floatstyle{ruled}
\newfloat{listing}{tb}{lst}{}
\floatname{listing}{Listing}

\usepackage{ulem}
\title{An Exploration into the Performance of Unsupervised Cross-Task Speech Representations for ``In the Wild'' Edge Applications}
\author {
    Heitor Guimarães$^1$, Arthur Pimentel$^1$, Anderson Avila$^2$, Mehdi Rezagholizadeh$^2$,
    Tiago H. Falk$^1$ }
\affiliations {
   $^1$ INRS-EMT, Université du Québec, Montreal, Canada\\
   $^2$ Huawei Noah’s Ark Lab\thanks{Any opinions, findings, and conclusions expressed in this manuscript are those of the authors and do not necessarily reflect the views, official policy or position of Huawei.}, Montreal, Canada\\
}

\begin{document}

\maketitle

\begin{abstract}
Unsupervised speech models are becoming ubiquitous in the speech and machine learning communities. Upstream models are responsible for learning meaningful representations from raw audio. Later, these representations serve as input to downstream models to solve a number of tasks, such as keyword spotting or emotion recognition. As edge speech applications start to emerge, it is important to gauge how robust these cross-task representations are on edge devices with limited resources and different noise levels. To this end, in this study we evaluate the robustness of four different versions of HuBERT, namely: base, large, and extra-large versions, as well as a recent version termed Robust-HuBERT. Tests are conducted under different additive and convolutive noise conditions for three downstream tasks: keyword spotting, intent classification, and emotion recognition. Our results show that while larger models can provide some important robustness to environmental factors, they may not be applicable to edge applications. Smaller models, on the other hand, showed substantial accuracy drops in noisy conditions, especially in the presence of room reverberation. These findings suggest that cross-task speech representations are not yet ready for edge applications and innovations are still needed.


\end{abstract}

\section{Introduction}
Learning to extract meaningful features from high-dimensional data is one of the main challenges in machine learning. For speech applications, extracting meaningful features that can be useful across tasks has been the focus of recent research (e.g.,  \citet{chen2022wavlm}). In fact, these learned representations will be crucial for the development of edge speech applications, where resource constraints will make it difficult or impossible to store and run different models on edge devices themselves.

Today, HuBERT, or hidden-unit bidirectional encoder representations from Transformers, exemplifies one of the leading self-supervised speech representations \cite{hsu2021hubert}. It has been widely used in the Speech Processing Universal Performance Benchmark (SUPERB), a benchmark of ten speech processing tasks built on established public datasets \cite{yang2021superb}. HuBERT has been made available to the community in different sizes, ranging from 95 million parameters in the base model, to 1 nearly billion parameters in the extra-large version. While its performance was recently shown to degrade in the presence of noise, a more robust version was recently proposed by \citet{huang2022improving} and referred to as here as Robust-HuBERT.

In this study, we are interested in better understanding of the impacts that additive and convolutive (room reverberation) noises might have on cross-task performance based on HuBERT representations. Moreover, since our ultimate goal is to apply these tasks on edge devices, it is important to gauge the impact of model size on overall accuracy. Here, our focus will be only on three tasks covering (i) a \textbf{content task} (keyword spotting), (ii) a  \textbf{semantic task} (intent classification), and (iii) a \textbf{para-linguistics task} (emotion recognition).

\section{Methods and Materials}

\begin{table*}[t]
    \centering
    \begin{tabular}{|cc||ccc|ccc|ccc|}
        \hline
        & & \multicolumn{3}{|c|}{Keyword Spotting} & \multicolumn{3}{|c|}{Intent Classification} & \multicolumn{3}{|c|}{Emotion Recognition} \\
        Upstream & \#params (M) & (c) & (n) & (n + r) & (c) & (n) & (n + r) & (c) & (n) & (n + r)  \\ 
        \hline\hline
        HuBERT Base & 95 & 0.9630 & 0.8377 & 0.5294 & 0.9831 & 0.7962 & 0.5663 & 0.6487 & 0.5272 & 0.3173 \\
        HuBERT Large & 317 & 0.9529 & 0.7478 & 0.4450 & 0.9873 & 0.8408 & \textbf{0.7171} & \textbf{0.6745} & 0.5696 & 0.4114 \\
        HuBERT X-Large & 964 & \textbf{0.9705} & 0.8747 & \textbf{0.7485} & \textbf{0.9921} & 0.8165 & 0.6599 & 0.6650 & 0.5731 & \textbf{0.4880} \\
        Robust HuBERT & 95 & 0.9662 & \textbf{0.9140} & 0.6702 & 0.9808 & \textbf{0.9262} & 0.7142 & 0.6470 & \textbf{0.5872} & 0.3494 \\ \hline
    \end{tabular}
    \caption{Experimental results for keyword spotting, intent classification, and emotion recognition under clean (\textbf{c}), noisy (\textbf{n}), and noise-plus-reveberation (\textbf{n+r}) test conditions.}
    \label{tab:results} \vspace{-4mm}
\end{table*}

\subsection{Self-supervised Speech Representations}
As mentioned previously, HuBERT is a self-supervised learning approach that attains representations based on speech-only data. The model relies on learning so-called hidden sound units (i.e. phonetic units), which can be used as frame-level targets \cite{hsu2021hubert}. Therefore, given an utterance $X = [x_1, x_2, ..., x_N]$, with $N$ acoustic frames, and the corresponding sequence of hidden unit targets $Z = [z_1, z_2 ,..., z_N]$, a clustering model must learn to assign $z_t$ to a set of classes $C = [c_1, c_2,..., c_K]$, where $C$ is a K-categorical variable. Hence, an offline clustering method is adopted to provide aligned target labels for a BERT-like prediction loss. This loss is applied over masked regions, forcing the model to learn combined acoustic and language model over continuous inputs, i.e.,:
\begin{equation}
    L_m(f; X, M, Z) = \sum_{t\in M} log p_{f}(z_t|\tilde{X},t),
\end{equation}
where $M \in [N]$ is the set of indexes to be masked in a sequence X of length $N$, and $f$ is a masked prediction model. Thus, the perturbed version of $X$ is represented by $\tilde{X} = r(X,M)$, where $r$ is a masking function. 


Note that the HuBERT model relies primarily on an unsupervised clustering step rather than on the quality of the clustering assignment labels. Starting with a simple k-means teacher of 100 clusters, the HuBERT model either matches or improves upon the wav2vec 2.0 model. The reader is referred to \citet{hsu2021hubert} for more details about the model.

Three variants of HuBERT are explored here. In the base version, the Transformer network has 12 layers. This number is doubled in the large version and quadrupled in the extra-large version. HuBERT-base is trained on 960 hours of the Librispeech dataset~\cite{panayotov2015librispeech}, whereas the large and extra-large versions are trained on the 60k hours from the Librilight dataset~\cite{kahn2020libri}. Robust-HuBERT, in turn, is a recent innovation that utilizes domain adversarial training, where the discriminator is a domain network responsible for classifying the source of distortion applied to the utterance, to improve accuracy in noisy conditions \cite{huang2022improving}.





\subsection{Downstream Tasks and Figure-of-Merit}
Three downstream tasks are used in our experiments. Keyword spotting is a 12-class classification problem where twelve speech commands need to be recognized. As this is an essential task for edge applications,  the robustness to real-world environments and model size are critical factors. Next, intent classification is used to assign labels to one of the three possible classes: action, object, and location. Lastly, emotion recognition, as the name suggests, is a multi-class problem where the emotion of the speaker is to be detected as either neutral, happy, sad, or angry. For all tasks, the evaluation metric is the recognition accuracy score, thus higher is better. For brevity, the interested reader is referred to \citet{yang2021superb} for more details on the tasks and the public datasets they rely on.

\subsection{Experimental Setup}
To simulate conditions to be observed in edge applications, additive and convolutive (i.e., room reverberation) noises are added to the SUPERB test sets. Here, the Deep Noise Suppression Challenge 4 noise dataset was used for this purpose \cite{dubey2022icassp}. The dataset is comprised of 180 hours of noise spread across 62,000 utterances covering 150 different noise types. To further simulate room reverberation, the openSLR28 dataset with 248 real room impulse responses and 60,000 synthetic ones from openSLR26~\cite{ko2017study} are used covering small-, medium-, and large-sized rooms. All files are resampled to 16~kHz. 

In our experiments, three conditions will be utilized: (1) ``clean (c)'' where test conditions as in SUPERB are used; (2) ``noisy (n)'' where the noises were added to the SUPERB test files at varying signal-to-noise ratios (SNR) uniformly sampled within the interval $[0, 20]$ dB; and (3) ``noise-plus-reverberation (n+r)'' where the room impulse responses are convolved with the SUPERB test files and then corrupted by additive noise, as per condition ``(n)''. To ensure that the same degradation conditions are present across all three tasks, a custom seed is adopted in the sampling process.

\section{Results}
Table~\ref{tab:results} shows the results obtained with our analysis for all three tasks and all three conditions. As it can be seen, for keyword spotting, accuracy increased when larger versions of HuBERT were explored. Robust-HuBERT achieved results in-line with HuBERT extra-large in the clean condition while requiring an order of magnitude fewer parameters. In fact, under the noisy condition, the robust version outperformed HuBERT extra-large, thus corroborating the advantages of adversarial training. Overall, noise and reverberation showed to be the most detrimental to keyword spotting, substantially reducing accuracy even for Robust-HuBERT. 


For intent classification, similar behaviour could be seen in clean conditions, where accuracy increased as model size increased. Robust-HuBERT achieved competitive performance in the noisy and noise-plus-reverberation conditions, outperforming even the extra-large model. The final accuracy, however, was substantially lower than that achieved in clean conditions, thus suggesting that there is still ample room for improvement for edge applications. Lastly, for emotion recognition, Robust-HuBERT showed to achieve the best accuracy in the noisy condition, but only sloightly better than the extra-large HuBERT model. The noise-plus-reverberation condition, again, showed to be the most challenging, resulting in accuracy dropping to almost half of what could be achieved with Robust-HuBERT in the clean condition. Overall, while extra-large models showed some resilience to noisy conditions, Robust-HuBERT achieved the best accuracy under these conditions. When noise and reverberation were present, however, all model performances degraded substantially.
\section{Conclusions}

In this work, we evaluate the performance of four variants of the HuBERT cross-task speech representation on three SUPERB tasks, across noisy and noise-plus-reverberation conditions. While larger models tend to achieve improved accuracy in clean conditions, performance degrades in noisy settings, thus suggesting that innovations are still needed, especially for edge speech applications where model size is an important limiting parameter.



\bibliography{aaai22}

\end{document}